\documentstyle[prl,aps,epsfig]{revtex}
\begin{document}
\draft \twocolumn[\hsize\textwidth\columnwidth\hsize\csname@twocolumnfalse\endcsname
\title{Anomalous Dynamic Scaling in Locally-Conserved Coarsening of Fractal Clusters}
\author{Azi Lipshtat and Baruch Meerson}
\address{The Racah Institute  of  Physics, Hebrew
University   of  Jerusalem, Jerusalem 91904, Israel}
\author{Pavel V. Sasorov}
\address{Institute of Theoretical and Experimental
Physics, Moscow 117259, Russia} \maketitle
\begin{abstract}
We report two-dimensional phase-field simulations of locally-conserved coarsening dynamics of
random fractal clusters with fractal dimension $D=1.7$ and $1.5$. The correlation function,
cluster perimeter and solute mass are measured as functions of time. Analyzing the correlation
function dynamics, we identify \textit{two} different time-dependent length scales that exhibit
power laws in time. The exponents of these power laws are independent of $D$; one of them is
apparently the ``classical" exponent $1/3$. The solute mass versus time exhibits dynamic scaling
with a $D$-dependent exponent, in agreement with a simple scaling theory.
\end{abstract}
 \vskip1pc] \narrowtext

Many non-equilibrium systems develop morphological instabilities and ramified growth at an early
stage of the dynamics, exhibit coarsening at an intermediate stage, and finally approach a simple
equilibrium. A classic example is diffusion-controlled systems, such as deposition of solute from a
supersaturated solution and solidification from an overcooled melt. The stage of morphological
instability has been under extensive investigation \cite{Langer1,Kessler,BMT,fractals}. If some
noise is present, fractal clusters (FCs) similar to diffusion-limited aggregates (DLA) can develop
\cite{BMT}. If the total amount of mass or heat is finite, the subsequent dynamics are dominated
by surface-tension-driven relaxation (coarsening) \cite{Irisawa,CMS,Kalinin}. Coarsening of FCs in
systems with conserved order parameter, apart from being interesting in its own right, exemplifies
a more general problem of \textit{phase ordering}: emergence of order from disorder, following a
quench from a disordered state into a region of phase coexistence \cite{Gunton,Bray}. This example
is non-trivial because of the long-ranged, power-law correlations intrinsic in FCs. The role of
long-ranged correlation in phase ordering dynamics has been under debate in connection to
\textit{dynamic scale invariance} (DSI), a major simplifying factor in theory \cite{Bray}. DSI
presumes that there is, at late times, a \textit{single} dynamic length scale $l(t)$ so that the
correlation function $C(r,t)$ approaches a self-similar form $g[r/l(t)]$ \cite{Bray}. In
locally-conserved systems (model B) $l(t)$ is expected to show dynamic scaling (by which we simply
mean a power law in time) with ``classical" dynamic exponent $1/3$ \cite{Gunton,Bray,LS}.

Because of complexity of phase-ordering dynamics, DSI has not been proven, except for a very few
simple models \cite{Bray}. However, there is extensive evidence, from experiments and simulations,
supporting DSI in conserved systems with \textit{short-ranged} correlations. Recently, phase-field
simulations of coarsening of a \textit{globally}-conserved, interface-controlled system with
\textit{long-ranged} correlations were performed \cite{PCM}. DLAs with fractal dimension $D=1.75$
served in Ref. \onlinecite{PCM} as the initial conditions for the minority phase. Notice that a FC
is a particular case of systems with long-ranged correlations. $D$ is equal to the exponent
$\sigma$ that appears in the power-law decay of the correlation function at $t=0$: $C(r,t=0) \sim
r^{-(d-\sigma)}$, where $d$ is the Euclidian dimension \cite{fractals}. The simulations \cite{PCM}
have not found any deviations from DSI and yielded the ``classical" value of the dynamic exponent
for the globally-conserved model, which is $1/2$.

The results \onlinecite{PCM} stand in sharp contrast with simulation results on
\textit{locally}-conserved coarsening of DLA clusters, where breakdown of DSI  was observed
\cite{CMS}. Breakdown of DSI is related to the fact that the upper cutoff $L$ of the FCs remains
almost constant in the process of coarsening \cite{Irisawa,CMS,Kalinin}. While interpreting this
finding, one should distinguish between two regimes: diffusion-controlled: $l_d < L$, and
Laplacian: $l_d > L$, where $l_d \sim t^{1/2}$ is the diffusion length. It is clear that $L$
should remain almost constant in the diffusion-controlled regime, as interaction between far-lying
parts of the cluster is exponentially small in this case. The inequality $l_d < L$ was satisfied in
the simulations of Refs. \onlinecite{CMS,Kalinin}, and it is satisfied in the simulations reported
in the present work. However, even in the opposite limit of Laplacian coarsening one can expect $L$
to remain almost constant (and breakdown of DSI to persist), this time because of Laplacian
screening of transport. An additional non-trivial aspect of \textit{locally}-conserved coarsening
of DLA clusters is the following. Though not dynamically scale-invariant, this coarsening process
was shown to exhibit dynamic scaling (with ``unusual" exponent $0.21-0.22$)
\cite{Irisawa,CMS,Kalinin}. This implies hidden simplicity of a more complex nature than DSI.

We report phase-field simulations of locally-conserved coarsening of 3 sets of random FCs. The
clusters have fractal dimensions $D=1.7$ and $1.5$. Our motivation was to check whether the
``anomalous" scaling persists for clusters different from DLA, and to investigate its possible
dependence on $D$.  The simulations confirm breakdown of DSI in all cases. We identify
\textit{three} power laws in time. Two of them correspond to dynamic length scales found from the
dynamics of $C(r,t)$ at small and intermediate distances. The first length scale, with exponent
$0.21-0.22$, is the same as observed earlier. Surprisingly, it does not show any dependence on
$D$. The second length scale has exponent close to $1/3$, and we suggest a simple interpretation
for its appearance. The third dynamic exponent is found in the time dependence of the ``solute
mass", and it is $D$-dependent. We suggest a simple scaling theory for the second and third
exponents.

Here is a brief description of our simulation techniques. We employed a hierarchical algorithm
\cite{Thouy} to build random FCs with tunable fractal dimension.  The algorithm starts with a set
of $2^n$ square particles grouped into pairs. These pairs are then grouped into pairs of pairs, and
so on, in an iterative procedure. After $n$ iterations the final aggregate of $2^n$ particles is
obtained. The desired fractal dimension $D$ is achieved by an appropriate mutual position in which
the aggregates are sticked together in each iteration \cite{Thouy}. The clusters obtained in this
way are "reinforced",  by an addition of peripheral sites as suggested in Ref.
\onlinecite{Irisawa}, to avoid breakup at an early stage of coarsening. The quality of obtained
FCs is controlled by computing the correlation function (see below) and checking the quality of the
power law.

A standard phase-field model for locally-conserved coarsening is the Cahn-Hilliard equation
\cite{Gunton,Bray}
\begin{equation}
\frac{\partial u}{\partial t} + \frac{1}{2}\nabla^2 \left(\nabla^2 u + u - u^3 \right) = 0 \,.
\label{CH}
\end{equation}
This equation was discretized and solved numerically. Two different numerical schemes were employed
to advance the solution in time. The first scheme used an explicit Euler integration. The
simulated domain $\Omega$ was a two-dimensional box 1024$\times$1024 with no-flux boundary
conditions (zero normal component of $\nabla u$ at the boundaries). The grid size was $\Delta
x=\Delta y=1$, the time step $\Delta t=1/25$ ensured numerical stability. The time range of these
simulations was $0<t<5000$. The second scheme used a semi-implicit Fourier spectral method
\cite{Zhu}. The simulation box was 1792$\times$1792, with periodic boundary conditions. The grid
size was $\Delta x=\Delta y=1$ that corresponds to $1792$ Fourier modes in each dimension. The
time step was $\Delta t=0.3$, the total time range was $0<t<15000$. Overall, three series of
simulations were performed. In two of them, with the explicit scheme, 10 FCs with $D=1.7$ (series
A) and 10 FCs with $D=1.5$ (series C) served as the initial conditions for the minority phase
$u=1$. In addition, 7 FCs with $D=1.7$ (series B) were simulated with the spectral scheme. The
minority phase area fractions were $10.5 \%$ (series A), $7.5 \%$ (series B) and $5.8 \%$ (series
C). Notice that the initial cluster area in series B was almost twice as big as that in series A,
while $D$ was the same.

Eq. (\ref{CH}) with either no-flux, or periodic boundary conditions obey conservation laws:
$I_0=\int\int\limits_{\hspace{-1em}\Omega} u({\mathbf r},t)\, {\mathbf d r} = const $ and $
{\mathbf I_1}=\int\int\limits_{\hspace{-1em}\Omega}{\mathbf r}\, u({\mathbf r},t)\, {\mathbf d r}
= const.$ Our numerical schemes preserve $I_0$ exactly. We checked that they also preserve
$\mathbf{I_1}$ with a very high accuracy: better than $5 \times 10^{-3}\,\%$ up to $t=5000$
(explicit scheme), and better than $0.05\,${\%} up to $t=15,000$ (spectral scheme). Additional
tests of the two schemes included obtaining the stationary kink solution of Eq. (\ref{CH}) and
observing the ``classic" scaling with exponent $1/3$ for the initial condition in the form of
``white noise".

\vspace{-0.1in}
\begin{figure}
%\center{\epsfxsize=5.9cm \epsffile{fig1.eps}} %\epsffile{fig1comp1.eps}}
\center{\epsfig{file=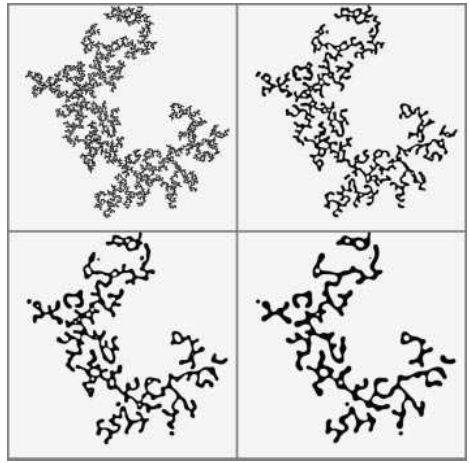, width=2.4in, clip= } }
\vspace{0.3cm}
\caption{Images of locally-conserved coarsening of a FC with $D=1.7$ at times $t=0$
(upper left), $495$ (upper right), $1963$ (lower left) and $4920$ (lower right). }
\end{figure}

\vspace{-0.3in}
\begin{figure}
%\center{\epsfxsize=5.9cm \epsffile{fig2.eps}} %\epsffile{fig2comp1.eps}}
\center{\epsfig{file=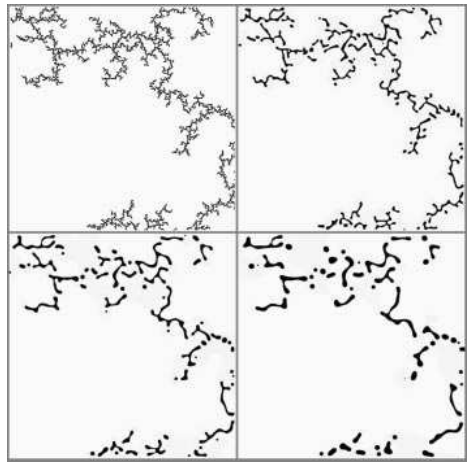, width=2.4in, clip= } }
\vspace{0.1cm}
\caption{Same as in Fig. 1, but for $D=1.5$.}
\end{figure}

The cluster was identified as the locus of $B(\textbf{r},t)=1$, where $B(\textbf{r},t)=1$ for
$u(\textbf{r},t) \ge 0$ and zero otherwise. Figures 1 and 2 show snapshots of the coarsening
dynamics observed in series A and C. One can see that larger features grow at the expense of
smaller ones. The global structure of the clusters remains unchanged. In particular, the cluster
size is almost constant, in a marked contrast to the globally-conserved case \cite{PCM}. A
significant decrease of $L$ in time is a necessary prerequisite of \textit{any} fractal coarsening
that obeys DSI \cite{CMS,PCM,Sempere,MS97}, therefore its absence clearly implies that DSI is
broken. One can also notice a more pronounced breakup of the cluster for $D=1.5$. Still, even in
this case the large-scale mass distribution in the cluster does not change much.

To characterize the coarsening dynamics, several quantities were sampled and averaged over the
initial conditions in each of the three simulation series:
\begin{enumerate}
\item (Circularly averaged)  equal-time correlation function, normalized at r=0:
$$
C(r,t)=\frac{<\rho({\bf r'}+{\bf r},t)\,\rho({\bf r'},t)>}{<\rho^{2}({\bf r'},t)>}\,.
$$
\item Cluster perimeter $P(t)$
\cite{imageproc}.
\item Cluster mass
$M(t)=\int\int\limits_{\hspace{-1em} B ({\mathbf r},t)=1}\rho\,({\mathbf r},t) \,\mathbf{d r}\,.$
\item ``Solute" mass outside the cluster \\
$M_s(t)=\int\int\limits_{\hspace{-1em} B ({\mathbf r},t)=0} \rho\,({\mathbf r},t) \,\mathbf{d
r}\,.$
\end{enumerate}
\noindent Here an auxiliary ``density" field $\rho({\bf r},t)=(1/2) [u({\bf r},t)+1]$ is
introduced that varies between $0$ and $1$. Also, $I_0=const$ implies that $M(t)+M_s(t)=const$.
\vspace{-0.1cm}
\begin{figure}
\center{\epsfxsize=7.0cm \epsfysize=4.7cm \epsffile{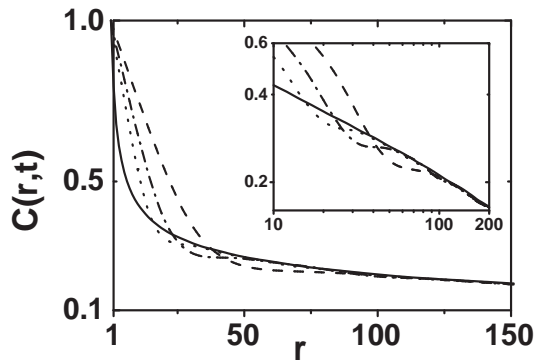}}
\vspace{0.2cm}\caption{Equal-time
pair correlation function at different times for simulation series B: $t=0$ (solid), $t=658.8$
(dotted), $t=2787$ (dashed-dotted) and $t=15000$ (dashed line). The inset shows a log-log plot of
the same function. The dynamic length scale $l_2(t)$ (the knee-like feature) is apparent.}
\label{Figcorr}
\end{figure}

Figure ~\ref{Figcorr} shows $C(r,t)$ at different times for simulation series B ($D=1.7$).
Coarsening occurs at small and intermediate distances, while the large-$r$ tail remains almost
unchanged. The fractal behavior is observed only in the ``frozen" tail that shrinks in time. On
the typical scales of coarsening, the cluster is not fractal anymore. This implies breakdown of
DSI at $D<2$. The same type of behavior of $C(r,t)$ was obtained for the locally-conserved
coarsening of DLA clusters \onlinecite{CMS,Kalinin}. At small distances, $C(r,t)$ behaves linearly
with $r$: $C(r,t)\simeq 1-r/l_1(t)$, as expected from the Porod law \cite{Bray}. This asymptotics
yields dynamic length scale $l_1 (t)$. One can also see a knee-like feature at intermediate
distances that develops in the course of time. This feature yields dynamic length scale $l_2$. We
define $l_2$ as the maximum value of $r$ for which the relative difference between $C(r,0)$ and
$C(r,t)$ is not greater than $3\%$ (the results are insensitive to the exact value of this
threshold). Notice that, by the end of the simulation time, the diffusion length $l_d$ is already
greater than $10^2$. Still, we do not see any signature of $l_d$ in the shape of $C(r,t)$ at large
distances. We expect that the tail of $C(r,t)$ will remain ``frozen" until late times, and DSI
broken, in the Laplacian regime as well.

Figures \ref{FigGraph17} and \ref{FigGraph15} show the measured quantities $l_1,\,l_2,\,P$ and
$M_s$, averaged over the initial conditions, versus time for simulation series B and C,
respectively. We fitted these time dependencies by power laws with exponents
$\alpha_1,\beta,\alpha_2$ and $\gamma$, respectively. According to Porod law \cite{Bray}, one
expects $\alpha_1=-\alpha_2$. All simulation results are summarized in Table 1. $\alpha_1$ and
$\alpha_2$ are close to the previously  reported values \cite{Irisawa,CMS,Kalinin}. Remarkably,
they do not depend on $D$.  The exponent $\beta$ is close to the classical exponent $1/3$. The
appearance of the classical exponent in a situation with broken DSI requires an explanation (see
below). The solute mass exhibits dynamic scaling with a $D$-dependent exponent $\gamma$.

Now we report a simple scaling theory for the exponents $\beta$ and $\gamma$. It is based on the
fact that, at subdiffusive distances, $r < l_d(t)$, the Cahn-Hilliard dynamics are reducible to a
sharp-interface model of Laplacian coarsening \cite{Bray,Pego}. The process of Laplacian
coarsening can be described as follows. Small branches of the cluster shrink and disappear, and
their material is reabsorbed by larger branches. Because of Laplacian screening, reabsorption
occurs locally (therefore, the mass distribution on large scales remains unchanged). The
reabsorption events cause undulations of the interfaces of the branches. Assuming DSI on length
scales $l_1(t) < r < l_d (t)$, we see that, by time $t$, the characteristic undulation lenght is
of order $t^{1/3}$. This gives a natural explanation to the dynamic length scale $l_2 \sim
t^{1/3}$ observed in the simulations. The Gibbs-Thomson condition at the interface
\cite{Bray,Pego} implies a solute density of order $t^{-1/3}$ in the ``contaminated" region within
a distance $l_d$ from the cluster. Taking into account the (preserved) fractal structure of the
cluster at distances large compared to $l_d$, we estimate the contaminated area as $A_d\sim
l_d^2\, (L/l_d)^D\propto t^{(2-D)/2}$. Multiplying the "solute density" $t^{-1/3}$ by $A_d$ yields
the solute mass: $M_s(t)\propto t^{(4-3D)/6}$. The last row in the table shows the theoretical
exponent $\gamma_{th} = (4-3 D)/6$. A good agreement with simulations is seen. Notice that
$\gamma_{th}$ changes sign at $D=4/3$. At $D<4/3$ reabsorption should effectively stop, and the
cluster should continue dissolving. We performed a single simulation with a random FC with D=1.3
that indeed showed a very slow but persistent increase of $M_s$ with time. Notice, however, that
FC with a smaller $D$ also implies, in our simulations, a smaller area fraction of the minority
phase. At too small area fractions the minority phase can dissolve completely, independently of
$D$. Therefore, a more careful investigation is needed in order to distinguish between these two
effects.
\begin{figure}
\begin{tabular}{cc}
  \epsfxsize=4.0cm  \epsffile{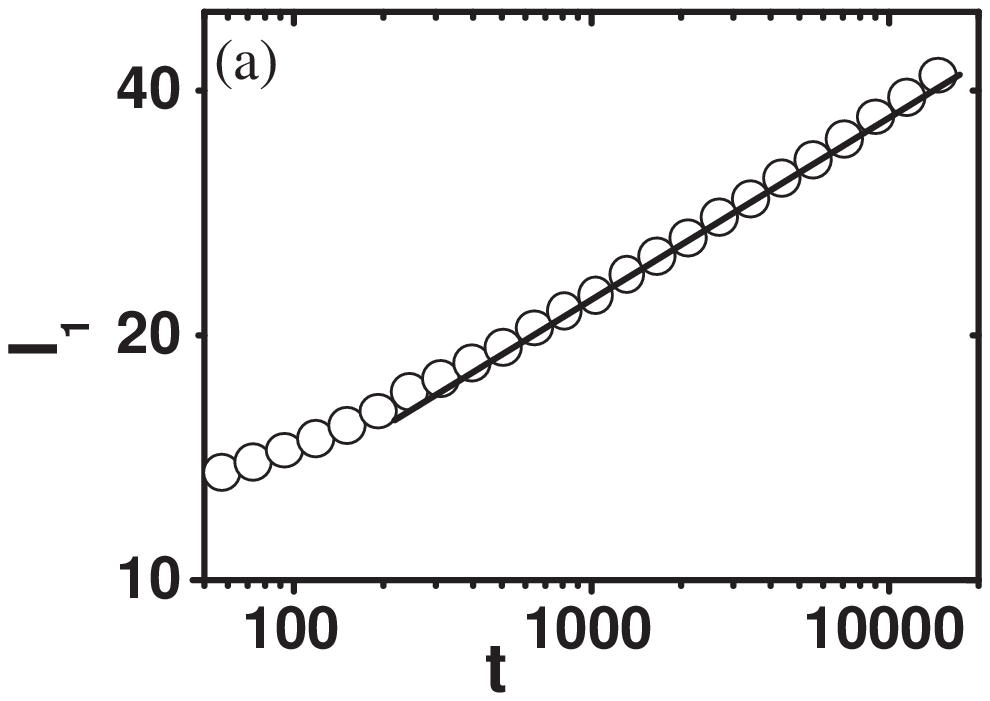} & \epsfxsize=4.0cm  \epsffile{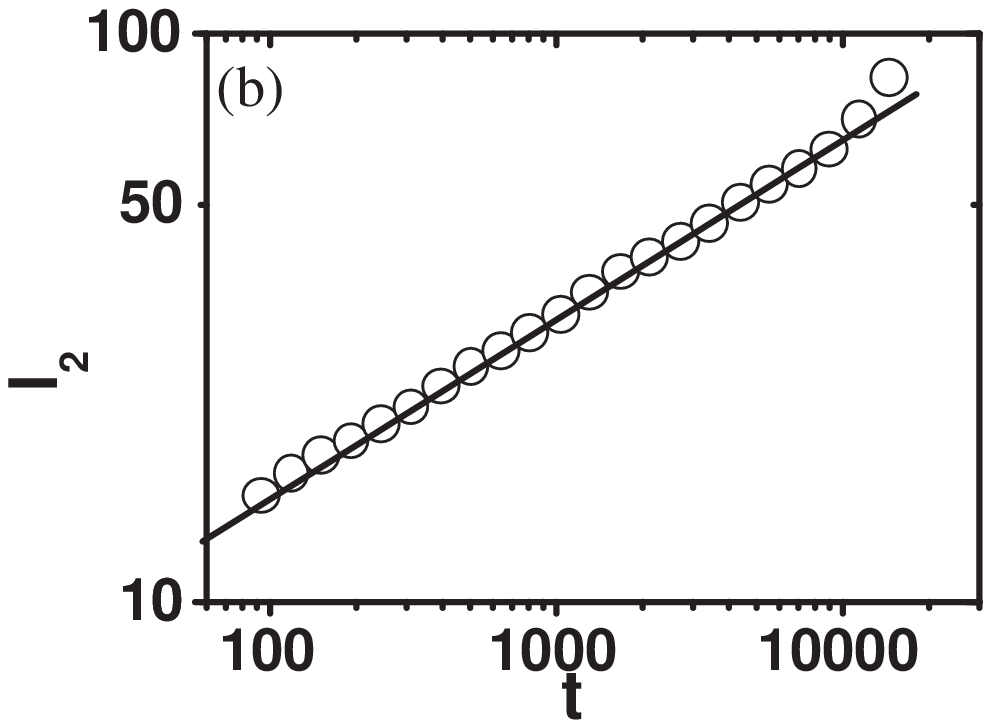}\\
  \epsfxsize=4.0cm  \epsffile{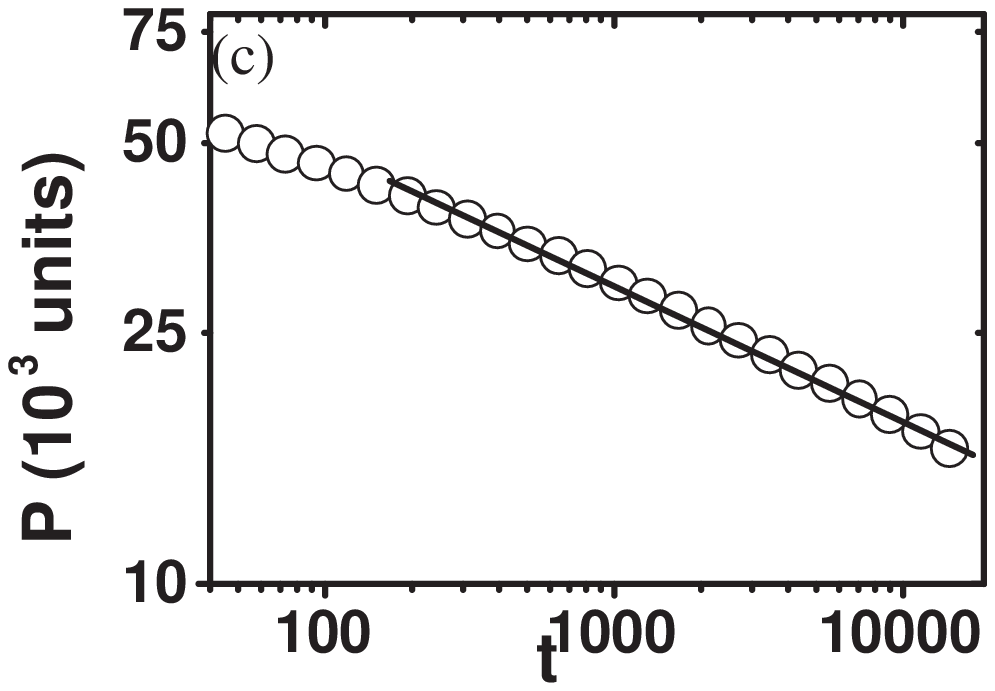} & \epsfxsize=4.0cm  \epsffile{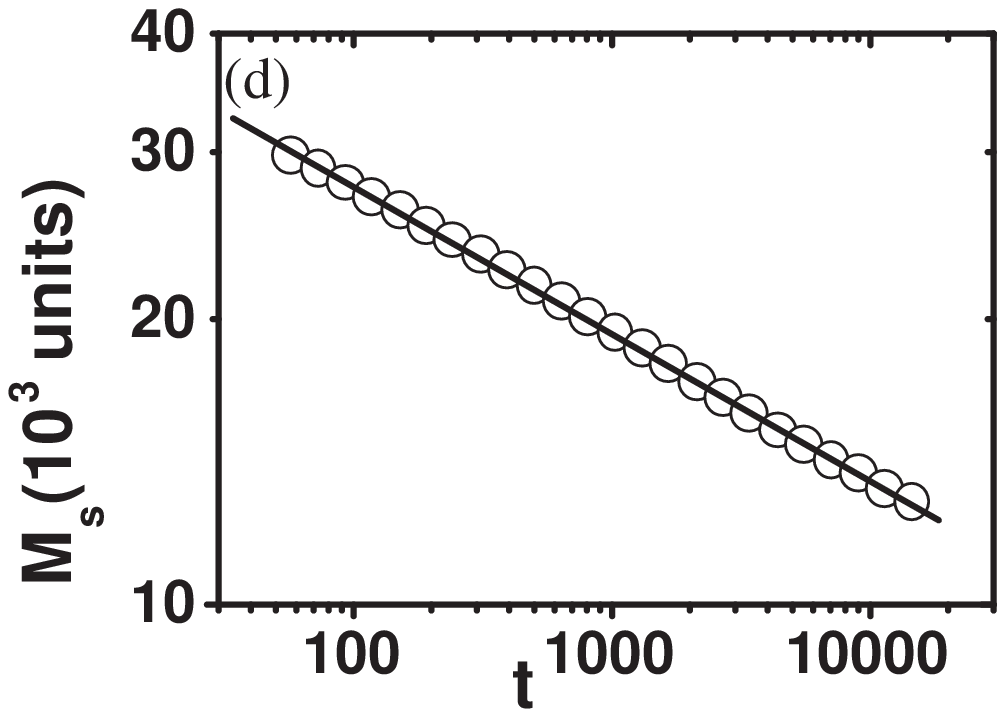}
\end{tabular}
\vspace{0.2cm} \caption{$l_1$ (a), $l_2$ (b), $P$ (c) and $M_s$ (d) vs. time for $D=1.7$. The solid
lines are power-law fits.} \label{FigGraph17}
\end{figure}
\vspace{-0.4cm}
\begin{figure}
\begin{tabular}{cc}
  \epsfxsize=4cm  \epsffile{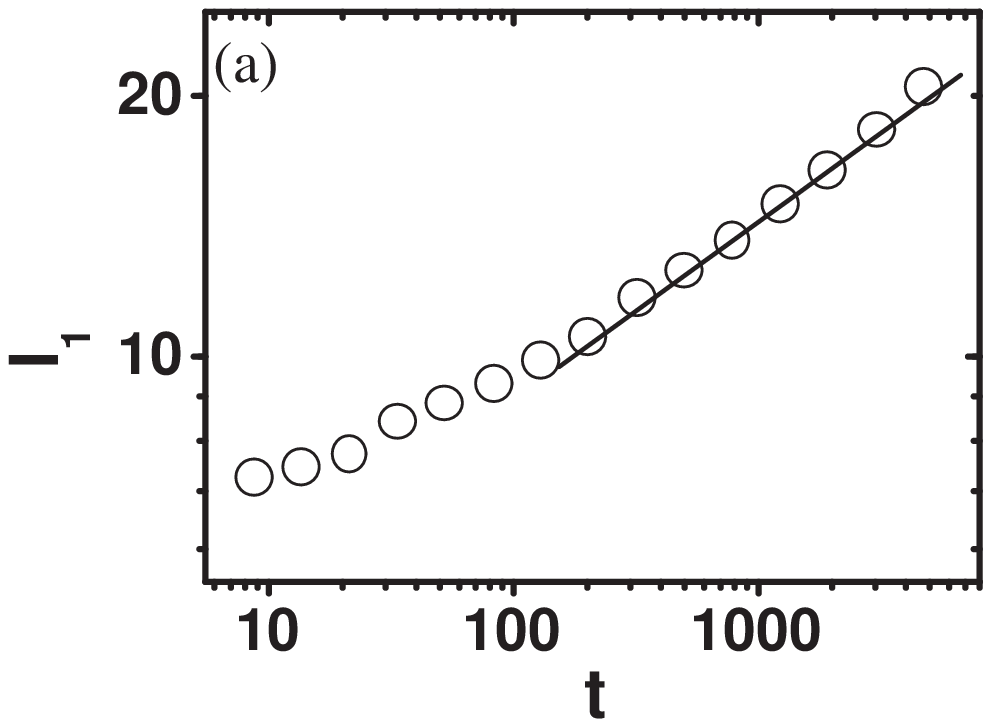} & \epsfxsize=4cm  \epsffile{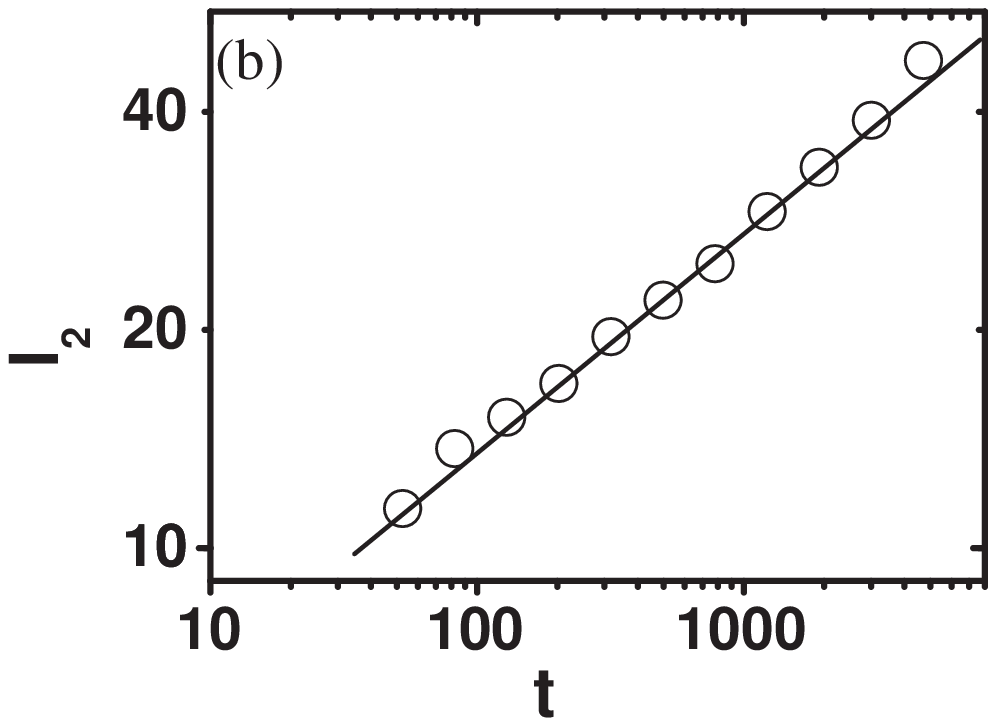}\\
  \epsfxsize=4cm  \epsffile{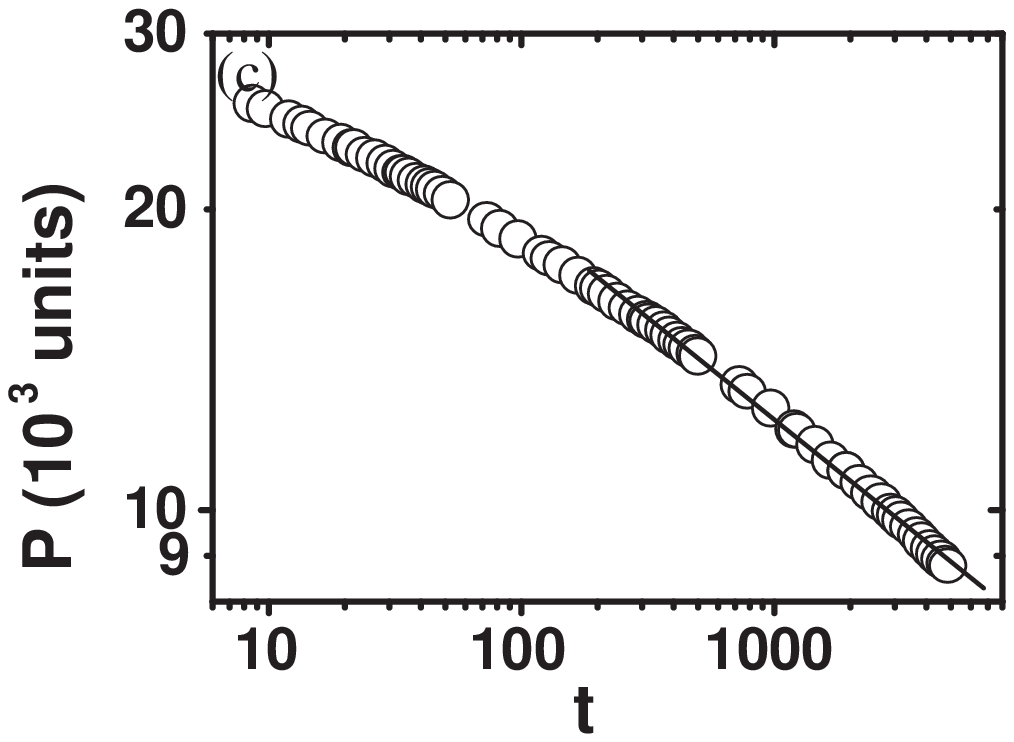} & \epsfxsize=4cm  \epsffile{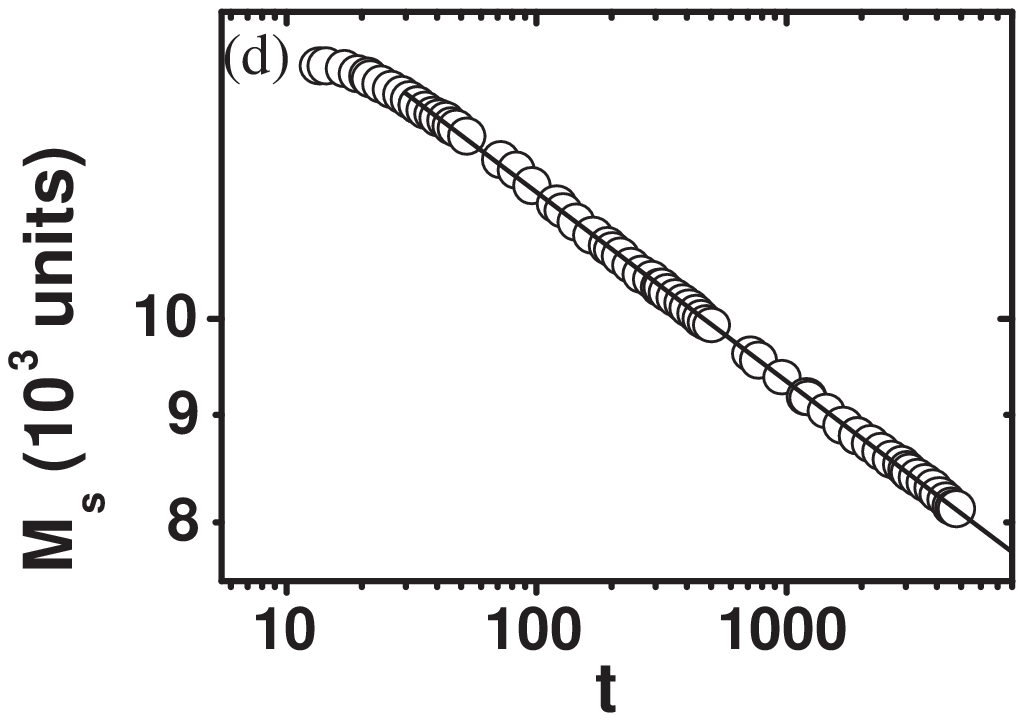}
\end{tabular}
\vspace{0.2cm}\caption{Same as in Fig. \ref{FigGraph17}, but for $D=1.5$.} \label{FigGraph15}
\end{figure}
\vspace{-0.3cm} \noindent
\begin{table}
\caption{Dynamic exponents found in the simulations. Also shown is $\gamma_{th}$, theoretical
prediction for $\gamma$. }
\begin{tabular}{|l||c|c|c|}
Series&A&B&C\\
$D$&$1.7$&$1.7$&$1.5$\\
\hline
$\alpha_1$&$0.21\pm0.01$&$0.22\pm 0.01$&$0.21\pm0.01$\\
$\alpha_2$&$-0.21\pm0.01$&$-0.21\pm 0.01$&$-0.21\pm 0.01$\\
$\beta$&$0.32\pm 0.01$&$0.32\pm 0.01$&$0.30\pm 0.01$\\
$\gamma$&$-0.16\pm 0.005$&$-0.16\pm
0.005$&$-0.09\pm 2\cdot10^{-4}$\\
\hline
$\gamma_{th}$&$-0.18$&$-0.18$&$-0.09$\\
\end{tabular}
{\footnotesize The errors are estimated by shifting the time intervals of fitting. The errors in
each fit are much smaller.}
\end{table} 

In summary, the simulations support earlier results \cite{CMS,Kalinin} on breakdown of DSI in
locally-conserved coarsening of systems with long-ranged correlations and reveal new signatures of
hidden simplicity in these systems. We confirm the value of the anomalous dynamic exponent
$\alpha_1 \simeq 0.21-0.22$. Surprisingly, it does not show any dependence on $D$. Two additional
dynamic exponents $\beta \simeq 1/3$ and $\gamma=\gamma (D)$ are identified. They can be
interpreted in terms of Laplacian coarsening at subdiffusive distances $r < l_d (t)$, combined
with DSI at intermediate distances $l_1(t) < r < l_d(t)$.

The work was supported by the Israel Science Foundation and by the Russian Foundation for Basic
Research (grant No. 99-01-00123). We thank Prof. M. Conti for providing the explicit Cahn-Hilliard
solver and correlation function diagnostics, and Nadav Katz for helping us with the realization of
the algorithm \cite{Thouy}. We also thank Prof. G.I. Barenblatt, Orly Lipshtat and Avner Peleg for
instructive discussions.

\end{document}